\documentclass[conference]{IEEEtran}
\IEEEoverridecommandlockouts
\ifCLASSOPTIONcompsoc
  \usepackage[nocompress]{cite}
\else
  \usepackage{cite}
\fi

\usepackage{amsmath,amssymb,amsfonts}
\usepackage{algorithmic}
\usepackage{textcomp}
\usepackage{xcolor}
\usepackage{booktabs}
\usepackage{graphicx,subfigure}
\usepackage{multirow}

\def\BibTeX{{\rm B\kern-.05em{\sc i\kern-.025em b}\kern-.08em
    T\kern-.1667em\lower.7ex\hbox{E}\kern-.125emX}}

\usepackage{fancyhdr}
    \makeatletter
    \newcommand{\linebreakand}{%
      \end{@IEEEauthorhalign}
      \hfill\mbox{}\par
      \mbox{}\hfill\begin{@IEEEauthorhalign}
    }
    \makeatother

\begin{document}

\title{Grounding Emotional Descriptions to Electrovibration Haptic Signals
\thanks{This work was supported by research grants from VILLUM FONDEN (VIL50296) and the National Science Foundation (\#2339707). 
}


\author{\IEEEauthorblockN{Guimin Hu}
\IEEEauthorblockA{
\textit{University of Copenhagen}\\
Copenhagen, Denmark \\
rice.hu.x@gmail.com}
\and
\IEEEauthorblockN{Zirui Zhao}
\IEEEauthorblockA{
\textit{Arizona State University}\\
Tempe, United States \\
ziruizhao0222@gmail.com}
\and
\IEEEauthorblockN{Lukas Heilmann}
\IEEEauthorblockA{
\textit{University of Copenhagen}\\
Copenhagen, Denmark\\
mail@lukasheilmann.com}
\linebreakand
\IEEEauthorblockN{Yasemin Vardar}
\IEEEauthorblockA{
\textit{Delft University of Technology}\\
Delft, Netherlands \\
y.vardar@tudelft.nl}
\and
\IEEEauthorblockN{Hasti Seifi}
\IEEEauthorblockA{
\textit{Arizona State University}\\
Tempe, United States\\
hasti.seifi@asu.edu}}
}

\maketitle
\thispagestyle{fancy}

\begin{abstract}
Designing and displaying haptic signals with sensory and emotional attributes can improve the user experience in various applications. Free-form user language provides rich sensory and emotional information for haptic design (e.g., ``This signal feels smooth and exciting''), but little work exists on linking user descriptions to haptic signals (i.e., language grounding). To address this gap, we conducted a study where 12 users described the feel of 32 signals perceived on a surface haptics (i.e., electrovibration) display. We developed a computational pipeline using natural language processing (NLP) techniques, such as GPT-3.5 Turbo and word embedding methods, to extract sensory and emotional keywords and group them into semantic clusters (i.e., concepts). We linked the keyword clusters to haptic signal features (e.g., pulse count) using correlation analysis. The proposed pipeline demonstrates the viability of a computational approach to analyzing haptic experiences. We discuss our 
future plans for creating a predictive model of haptic experience.
\end{abstract}


\section{Introduction}
Programmable touch signals (i.e., haptics) with sensory and emotional attributes can provide immediate feedback, enhancing user experience in various applications like gaming, virtual reality environments, and mobile applications. For example, a lively haptic signal can increase user enjoyment during game play~\cite{singhal2021juicy} or encourage the user to continue an action, while an alarming signal can warn the user about potential errors such as deleting a file~\cite{levesque2011enhancing}. Children can feel the emotional sentiment of words in a storybook with haptic effects. Also, haptics offers an alternative means of receiving emotional and contextual information for users with visual or auditory impairments.

Haptics researchers have proposed various methods to assess the emotional content of haptic signals \cite{seifi2013first,yoo2015emotional}.
Some works collected free-form user descriptions to showcase the rich range of sensations evoked by haptic signals \cite{obrist2013talking}. User descriptions provide insights into the sensory and emotional attributes of haptic signals, but computational approaches for analyzing user language are missing in haptics. For example, a user may describe an electrovibration signal as ``this signal feels smooth and exciting'' or 
``it's not super strong. It's really quite pleasing.'' 
Haptics researchers manually analyze these descriptions to extract relevant keywords and identify trends or themes across various users~\cite{clarke2021thematic}. Yet, the manual data analysis is time-consuming and difficult to scale. This difficulty is exacerbated since haptic designers often need to test various combinations of signal parameters in user studies to verify the sensory and emotional content of the signals~\cite{seifi2013first,schneider2017haptic}.
To accelerate haptic signal design, we need efficient methods for analyzing user descriptions and linking them to haptic signal features such as pulse count (i.e., language grounding).
\begin{figure}[t]
\centerline{\includegraphics[width=0.8\linewidth]{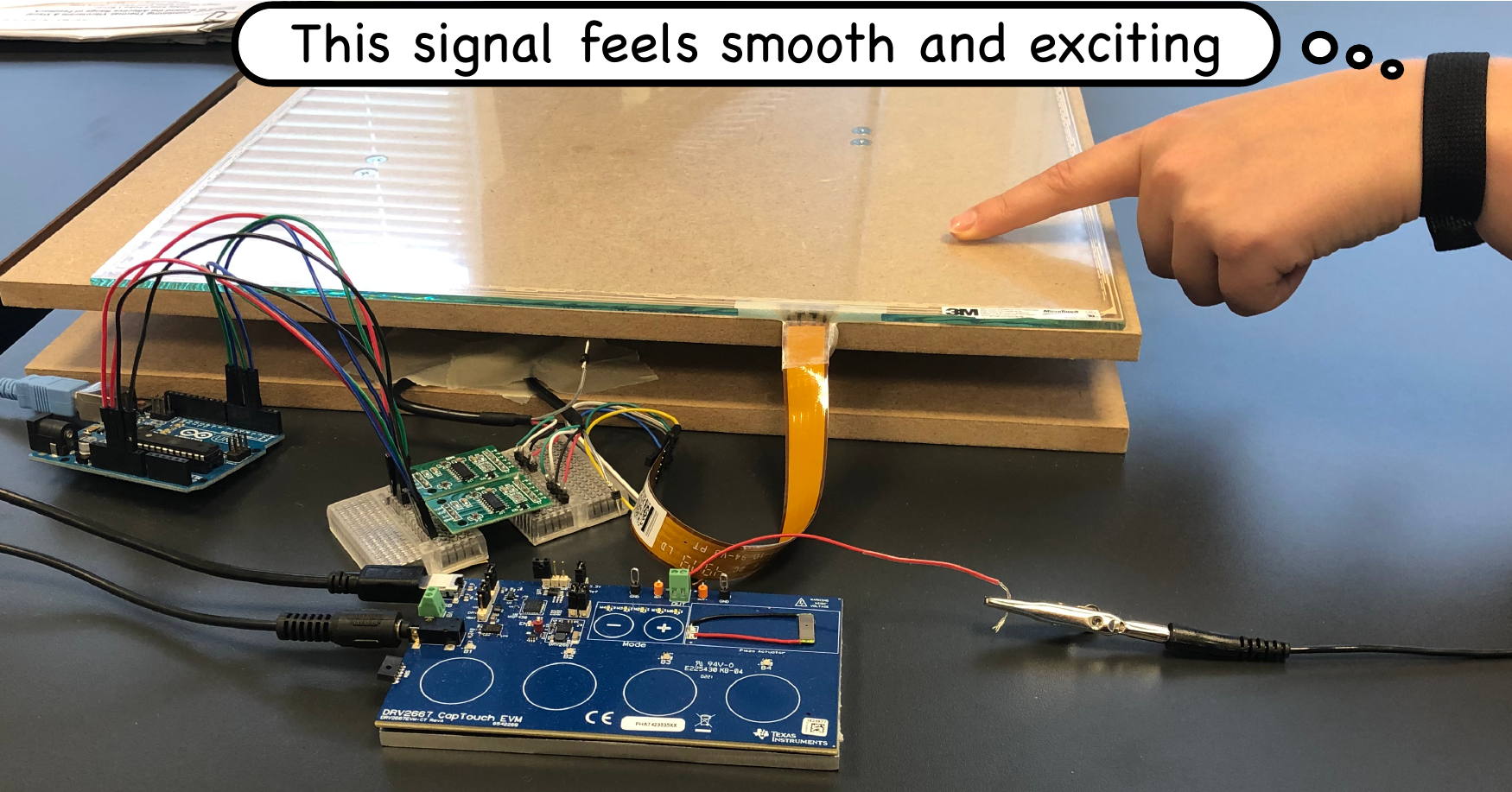}} 
\caption{The electrovibration haptic device in our data collection user study and an example of free-form user description.}
\label{fig:teaser}
\end{figure}

In this paper, we leveraged natural language processing (NLP) techniques to analyze free-form user descriptions 
for surface haptic signals~\cite{basdogan2020review,vardar2017effect}. First, we collected user descriptions for 32 surface electrovibration signals in a lab study with 12 users (16 signals per participant). During the study, the participant(s) listened to white noise on a headphone and felt each haptic signal by sliding their index finger over a 3M MicroTouch electrostatic screen without seeing any visual representation of the signal (Figure~\ref{fig:teaser}). They verbally described the sensory, emotional, and associative attributes of the tactile sensation. All sessions were audio-recorded and transcribed. 
Next, we developed an NLP pipeline to analyze the descriptions in three steps. 
First, we used NLP techniques to extract sensory and emotion keywords from free-form user descriptions. Second, we divided the keywords into positive and negative sets to form semantic concepts (i.e., keyword clusters). Finally, we linked the keyword clusters to haptic signal features (e.g., pulse count) using correlation analysis.  
We contribute:
\begin{itemize}
\item {The first computational pipeline for analyzing free-form haptic descriptions}
\item {Clusters of positive and negative keywords used for describing haptic experience}
\item {Preliminary results on linking the concept clusters and electrovibration signal features using correlation analysis}
\end{itemize}
Our late-breaking results suggest a promising direction for haptic researchers is to further explore computational approaches for haptic signal design through NLP techniques.

\begin{figure*}[t]
\centerline{\includegraphics[width=0.70\textwidth]{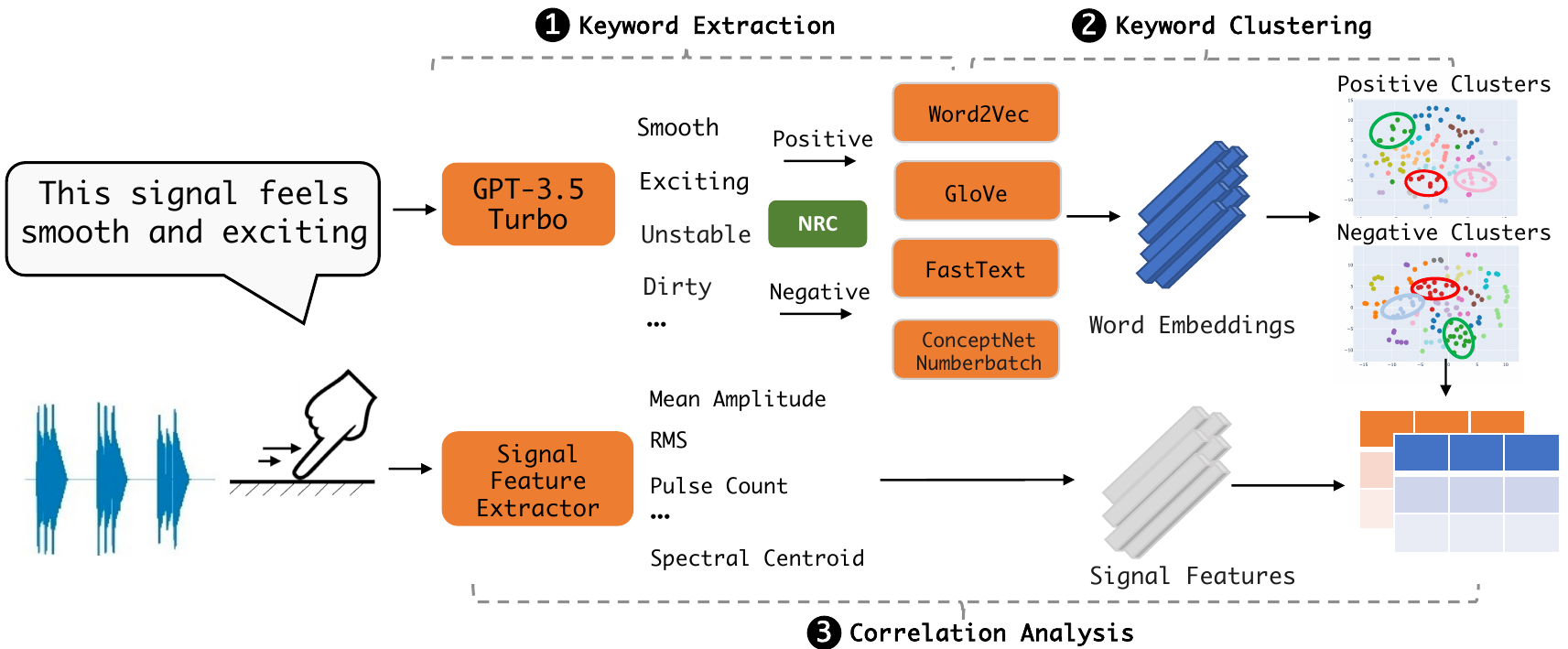}}
\caption{The overview of our computational pipeline with three main phases: (1) Keyword extraction identifies sensory and emotional keywords from the user descriptions and groups the keywords into positive and negative sentiments, (2) Keyword clustering clusters the keywords using four word embedding methods, and (3) Correlation analysis extracts statistical signal features and calculates the correlations between signal features and concept clusters.}
\label{fig:pipeline}
\end{figure*}

\section{Data Analysis and Results}
Our data analysis pipeline is composed of three main phases (Figure~\ref{fig:pipeline}): 
(1) \emph{keyword extraction} selects the sensory and emotional words from the user description of the haptic signal, 
(2) \emph{keyword clustering} identifies groups of keywords with similar semantics and is akin to the current practice of identifying concepts or \textit{themes} manually by haptics researchers~\cite{obrist2013talking}, and  
(3) \emph{correlation analysis} describes the link between haptic signal features (e.g., pulse count) and the concept clusters, to find the set of words related to the tactile experience.

\subsection{Keyword Extraction}
Most words in the verbal descriptions are unrelated to the subjective experience of the haptic signals. An example user description in our dataset is as follows: ``it feels quite \textit{smooth}, like putting your finger in \textit{water}. I do \textit{not} feel \textit{excited}. I would say it's \textit{boring}.'' We extracted the words marked in italics as the keywords about user experience. Rule-based regular expressions or Part-of-speech (POS) tagging can be used for keyword extraction. Yet, the rule-based approach is limited to preset sentence templates, and POS tagging can only extract single words (i.e., not phrases).
Thus, we adopted a prompt-based approach with large language models (LLM) for keyword extraction. The prompt template is as follows: ``Extract keywords including sensational, emotional, metaphoric, and usage examples from the corresponding texts below.'' We fed the prompt into GPT-3.5 Turbo to return the emotional and sensory keywords.
Table~\ref{tab:key_extraction} compares the performance of the above extraction methods against the manually-extracted keywords by one author using precision (P), recall (R), and F1 as the evaluation metrics. GPT-3.5 Turbo showed higher precision, recall, and F1 scores than rule-based and POS-based methods. Thus, we used the keywords extracted by GPT-3.5 Turbo for the subsequent steps. 

\begin{table}[!h]
    \centering
        \caption{Performance of the three keyword extraction methods.}
    \begin{tabular}{l|ccc}
    \toprule
    Keyword Extraction Method &{Precision}&{Recall}&{F1}\cr
      \midrule
      Rule-based&0.37&0.49&0.41\\
      POS-based&0.37&0.62&0.47\\
      GPT-3.5 Turbo (with fine-tuning)&0.71&0.70&0.70\\
      \bottomrule
    \end{tabular}
    \label{tab:key_extraction}
\end{table}


\begin{figure}[t]
\centering
\subfigure
{
\includegraphics[width=0.45\linewidth]{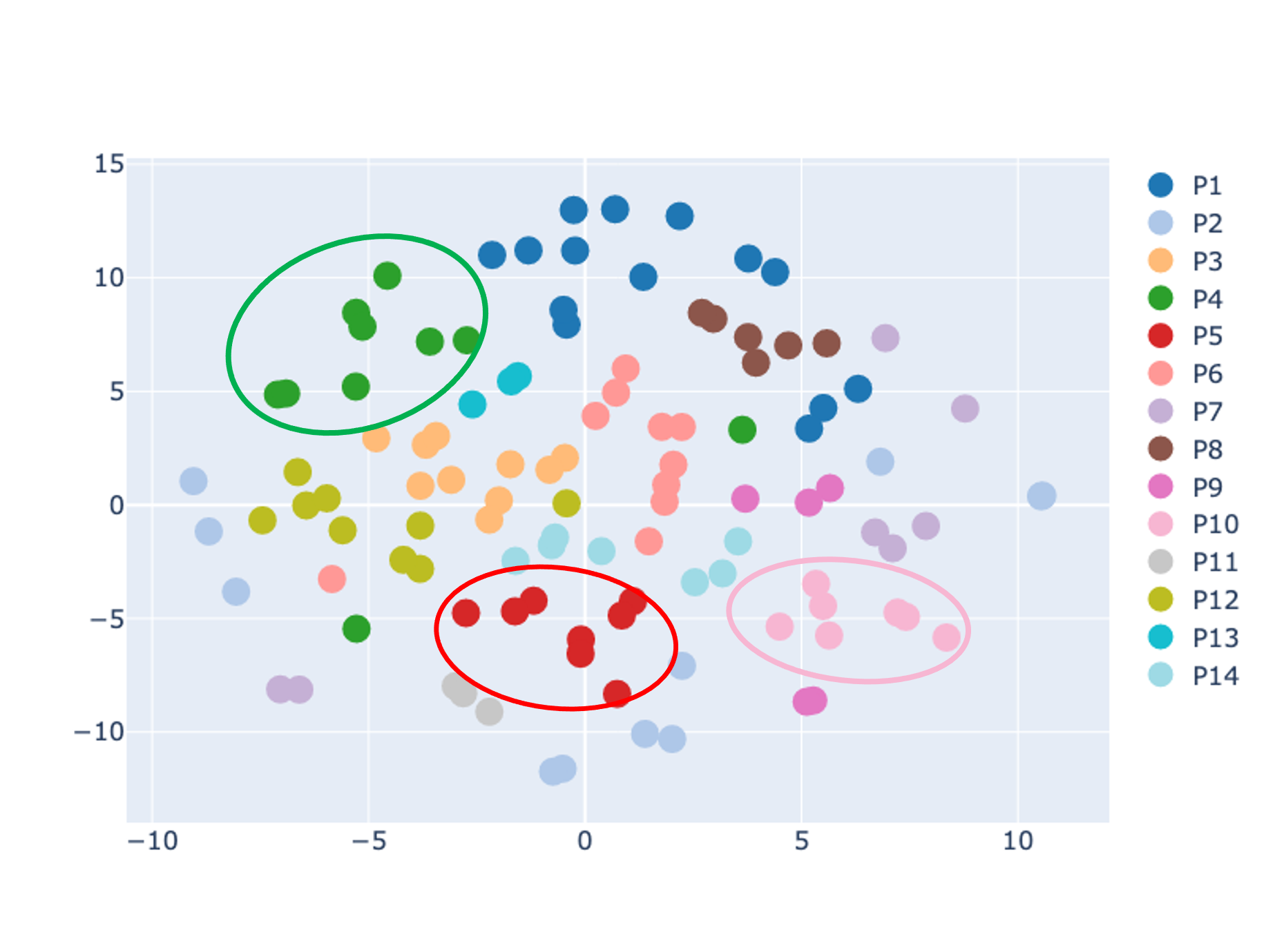}
}
\hspace{0.05in}
\subfigure
{
\includegraphics[width=0.45\linewidth]{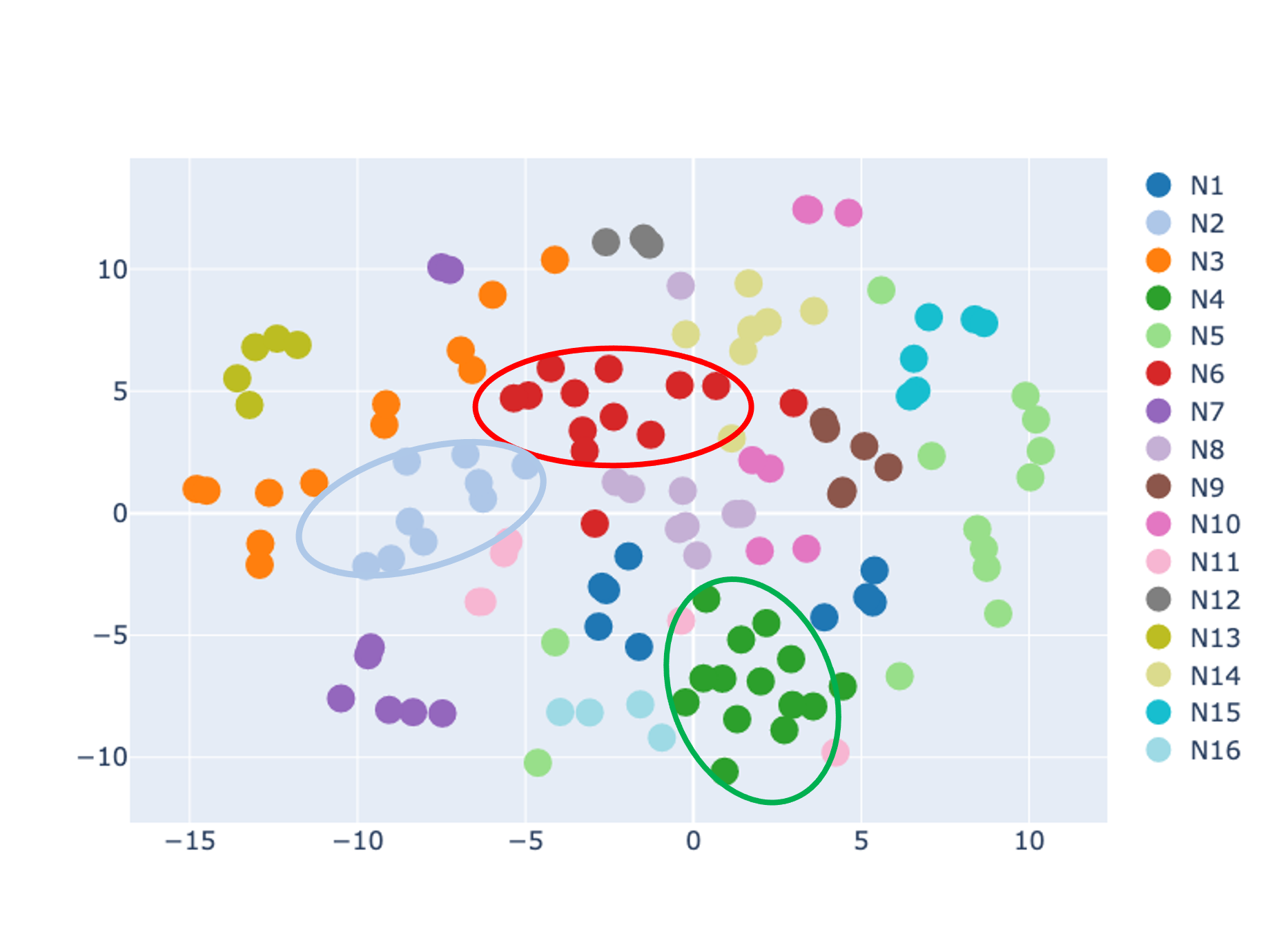}
}
\caption{T-SNE visualization of the ConceptNet Numberbatch word embeddings and clusters of positive (left) and negative (right) keywords in our dataset. Circles showcase example clusters to denote coherence in the results.}
\label{fig:cluster_visualization_cn}
\end{figure}


\begin{table*}[]
 \begin{minipage}{.5\linewidth}
    \caption{Top three frequent keywords for the 14 positive clusters.} 
    \centering  
\begin{tabular}{l|l}
\toprule
Clusters 1--7                   & Clusters 8--14                                  \\
\midrule
$P_{1}$: urgent, important, notification &  $P_{8}$: constant, steady, stable\\
$P_{2}$: massage, billowy, beach & $P_{9}$: rhythmical, impatient, staccato \\
$P_{3}$: quiet, soothing, calm  & $P_{10}$: sudden, surprised, abrupt            \\
$P_{4}$: passive, neutral, active & $P_{11}$: musical, creative, artistic            \\
$P_{5}$: funny, playful, silly  &  $P_{12}$: familiar, warm, tender       \\
$P_{6}$: satisfied, clean, confident & $P_{13}$: vigilant, attentive, discreet      \\
$P_{7}$: buzz, electric, sensationless & $P_{14}$: pleasant, stimulative, exciting   \\
\bottomrule
\end{tabular}
\label{tab:top_pos_keywords}
\end{minipage}
\begin{minipage}{.5\linewidth}
\scriptsize
    \centering
        \caption{Top three frequent keywords for the 16 negative clusters.} 
\begin{tabular}{l|l}
\toprule
Clusters 1--8                   & Clusters 9--16                                     \\
\midrule
$N_{1}$: unenthusiastic, numb, indifferent  & $N_{9}$: unpredictable, insecure, unstable\\
$N_{2}$: weird, mysterious, shady & $N_{10}$: urgent, difficult, imperative \\
$N_{3}$: small, billowy, bee  & $N_{11}$: startling, abrupt, sudden \\
$N_{4}$: uncomfortable, anxious, impatient & $N_{12}$: restrictive, limited, restriction  \\
$N_{5}$: lagging, assailable, distrust  & $N_{13}$: erase, forget, stumble \\
$N_{6}$: dirty, sticky, bad & $N_{14}$: aggressive, intense, passive \\
$N_{7}$: quiet, alarm, silent& $N_{15}$: frictional, resistive, resistant \\
$N_{8}$: annoying, irritating, oppressive & $N_{16}$:  silly, crazy, frantic   \\
\bottomrule
\end{tabular}
\label{tab:top_neg_keywords}
\end{minipage}
\end{table*}

\begin{table}[t]
    \centering
\caption{Largest cluster correlation for each signal feature. We include multiple correlations when they are close ($\le.02$ difference) and the most frequent keyword for each cluster. Plus and minus signs show the direction of correlations.}
\resizebox{0.95\linewidth}{!}{
\begin{tabular}{l|ll}
\toprule
Signal Features         & Positive Keyword Clusters &  Negative Keyword Clusters\\
\midrule
Mean Amplitude        & $-P_{11}$: musical & $+N_{9}$: unpredictable  \\
RMS                   & $+P_{1}$: urgent & $+N_3$: small \\
Pulse Count           & $-P_{8}$: constant & $+N_3$: small\\
Std Pulse Distance    & $-P_{2}$: massage & $-N_2$: weird, $-N_3$: small\\
Zero Count            & $+P_9$: rhythmical & $-N_3$: small\\
Mean Onset Strength   & $-P_8$: constant & $+N_2$: weird, $+N_3$: small\\
Spectral Centroid     & $-P_2$: massage & $-N_3$: small, $+N_{14}$: aggressive\\
\bottomrule
\end{tabular}}
\label{tab:corr}
\end{table}

\begin{figure*}[t]
\centering
\subfigure
{
\includegraphics[width=0.44\linewidth]{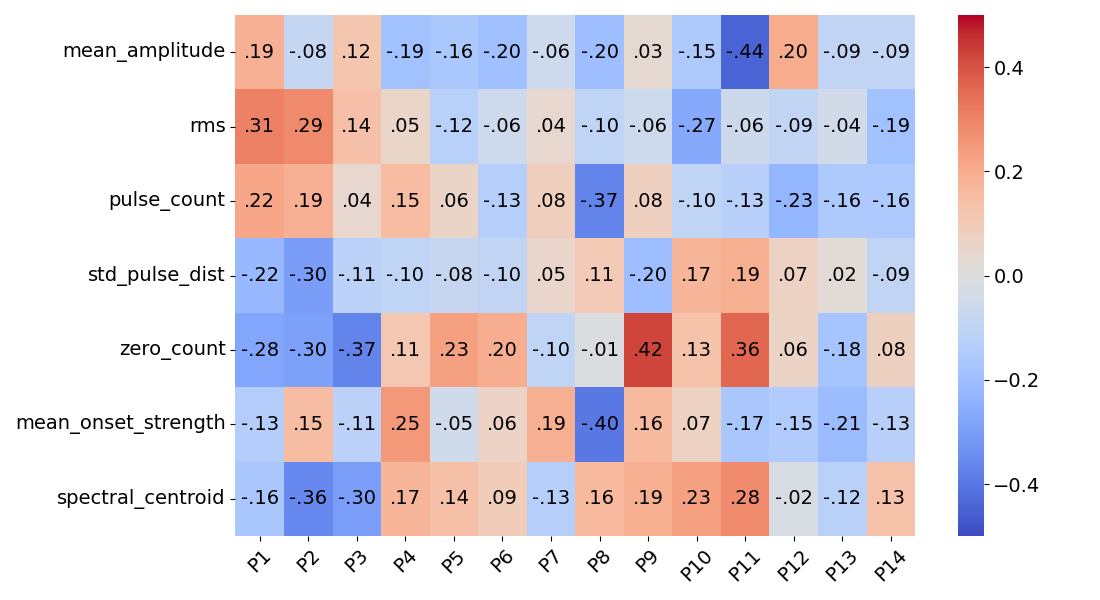}}
\hspace{0.05in}
\subfigure
{
\includegraphics[width=0.44\linewidth]{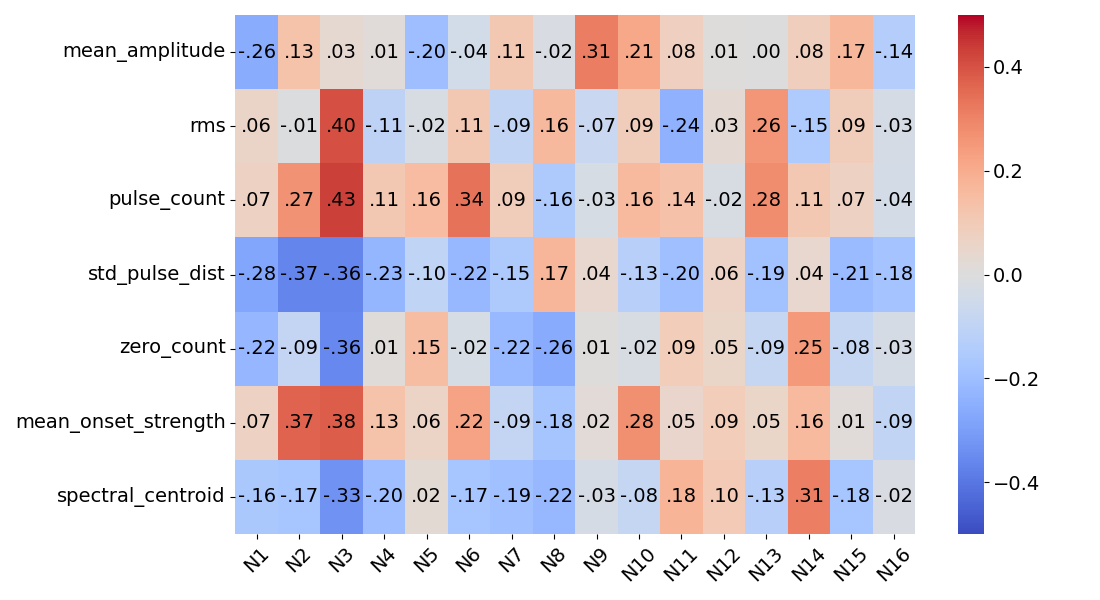}}
\caption{The correlations between signal features and the clusters for positive (left) and negative (right) keywords.}
\label{fig:corr_visualizations}
\end{figure*}
\subsection{Keyword Clustering}


To identify the emotional attributes of haptic signals, we divided the keywords into positive and negative sentiment groups using the NRC emotional lexicon~\cite{mohammad2013nrc}. 
Then, we used four pre-trained static word embeddings Word2Vec~\cite{DBLP:conf/nips/MikolovSCCD13}, 
FastText~\cite{joulin2017bag,joulin2016fasttext}, 
GloVe~\cite{DBLP:conf/emnlp/PenningtonSM14}, 
and ConceptNet Numberbatch~\cite{speer2017conceptnet} 
to project the keywords into vectors. These embeddings allowed us to quantify the semantic similarity among words by measuring the distance between their vectors. 
After obtaining the embeddings of keywords, we applied agglomerative hierarchical clustering~\cite{murtagh2012algorithms} to group the positive and negative keywords into semantic concepts. 
For example, warning, alarm, and siren can be in the same cluster, reflecting an urgent sensation. 


We compared the visualization and statistics of clusters from the four embedding methods to evaluate their suitability for identifying concepts related to the sensory and emotional haptic experience. 
Figure~\ref{fig:cluster_visualization_cn} shows the visualizations of clusters for positive and negative keywords using the ConceptNet Numberbatch. In the figure, the embeddings of the words in the same cluster are close in their representation space, suggesting that hierarchical clustering can capture concepts based on the semantic information of keywords. 
Thus, we used the word embedding from ConceptNet Numberbatch for the correlation analysis in the next step. Tables~\ref{tab:top_pos_keywords} and \ref{tab:top_neg_keywords} show the top three most frequent keywords for the clusters from ConceptNet Numberbatch (See Supplemental Materials for GloVe, Word2Vec, and FastText results).

\subsection{Correlation Analysis with Haptic Signal Features}
To assess whether the user keywords can be predicted from the signals, we extracted various signal features (e.g., mean amplitude) from the haptic signals and then calculated the correlation between the signal features and concept clusters. Specifically, we extracted and normalized seven signal features that reflect the characteristics of the electrovibration waveform based on prior literature on the perception of tactile signals~\cite{fishel2012bayesian, strese2020classification, richardson2022learn2feel}.  
For example, mean amplitude and root mean square (RMS) represent the signal's strength and energy, while the standard deviation (STD) of pulse distances measures the rhythmic regularity based on the duration of silences between consecutive pulses. 
We counted the number of keywords from each cluster in the descriptions for the 32 signals. 
This step resulted in a matrix of 32 signals$\times$14 positive clusters and a matrix of 32 signals$\times$16 negative clusters. Then, we calculated Pearson's correlations as a measure of the strength of the relation between signal features and concept clusters.  

Figure~\ref{fig:corr_visualizations} presents the correlation results, and Table~\ref{tab:corr} denotes the largest correlations between signal features and concept clusters, providing the most relevant concepts as the value of a feature increases or decreases. 
Specifically, keywords from many clusters (e.g., urgent, rhythmical, weird, constant) are shared with prior haptic literature. Also, several correlations are easy to interpret. For example, 
Higher RMS (i.e., signal energy) correlates with keywords such as urgent. Signals with higher pulse count and mean onset strength are perceived as less constant and steady. Signals with higher STD of pulse distance are irregular and feel less like a massage, or beach. Our results also revealed that a few concept clusters correlate with multiple signal features, and some clusters only showed weak correlations. For example, six features correlate with N3 (small), referring to a small sensation or buzz. On the other hand, N16 (silly) has weak correlations with all features. The first observation suggests that various electrovibration signals can induce the same sensory and emotional associations, providing designer flexibility. The latter suggests that some emotional associations may require more complex signals than others and that a combination of several features must be used to predict those feelings. Another reason could be that the selected signal features are not rich enough to capture these emotional descriptors. 

\section{Conclusion}

This work is the first to use a computational approach to analyze free-form user descriptions for haptics. The weak to moderate correlations in our results suggest that a subset of descriptions are shared across users and can be linked or grounded to signal features. 
However, our dataset is small and has limited diversity. Thus, the results may not generalize to a larger diverse set of signals and user groups. 
Our ongoing plans include collecting a larger dataset of haptic signals and user descriptions to strengthen the validity and generalizability of the findings. Then, we can develop haptic word embeddings with the pre-trained language models and we will also explore non-linear approaches such as multi-label classification for predicting haptic emotions.

\section*{Ethical impact statement}
Our work on analyzing sensory and emotional descriptions for haptic signals raises three main ethical considerations.
First, the emotional responses of users regarding haptic experience can be considered a private matter. 
Second, our dataset is limited to young English-speaking adults without sensory impairments and is not representative of the broader population. 
Third, the collection and handling of personal data in our study was approved by the university's ethics board. Also, we deleted all the recorded audio files after transcription.
\bibliography{IEEEtran} 
\bibliographystyle{IEEEtran}
\end{document}